\documentclass[rmp,letterpaper,twocolumn,superscriptaddress,floatfix]{revtex4}

\usepackage{graphicx}
\usepackage{natbib}

\begin{document}

\title{Self-Repairing Peer-to-Peer Networks}
\author{G\'abor Cs\'ardi}
\email[Corresponding author, ]{csardi@rmki.kfki.hu}
\affiliation{Department of Biophysics, KFKI
  Research Institute for Particle and Nuclear Physics of the
  Hungarian Academy of Sciences, Budapest, Hungary}
\affiliation{Kalamazoo College, Kalamazoo, MI, USA}
\author{Maxwell Young}
\affiliation{Computer Science Department at the University
  of New Mexico, NM, USA}
\author{Jennifer Sager}
\affiliation{Computer Science Department at the University
  of New Mexico, NM, USA}
\author{P\'eter H\'aga}
\affiliation{Department of Physics of Complex Systems,
  E\"otv\"os University, Budapest, Hungary}

\begin{abstract}
  In this paper we study the resilience of peer-to-peer networks to
  preferential attacks. We define a network model and experiment with
  three different simple repairing algorithms, out of which the so
  called `2nd neighbor' rewiring algorithm is found to be effective
  and plausible for keeping a large connected component in the
  network, in spite of the continuous attacks. While our motivation
  comes from peer-to-peer file sharing networks, we believe that our
  results are more general and applicable in a wide range of
  networks. 
  All this work was done as a student project in the Complex Systems
  Summer School 2004, organized by the Santa Fe Institute in Santa Fe,
  NM, USA. 
\end{abstract}

\maketitle


\section{Introduction}


There has been an increasing interest in the field of complex networks
recently. Describing complex systems as graphs has proven to be an
effective approach and it is now widely adopted. Network analysis and
modeling has contributed to understanding mechanisms and organization
principles of various systems \citep{newman03,albert02,dorogovtsev02}.


In computer science much attention has turned to \emph{peer-to-peer}
computer networks, sometimes abbreviated as P2P networks. In a
peer-to-peer computer network, every (node) computer has the same
capabilities and responsibilities, there are no distinguished
nodes. This is in contrast to client/server architectures, where the
roles (and usually also the capabilities) of the client and server
nodes differ.

Many peer-to-peer computer networks provide file-sharing
services (but not all of them, see the work of \citet{milojicic02}),
the owners of the network nodes intend to share their files 
with the other node owners in the same network. It is observed that
many nodes of these file sharing networks distribute illegal content,
most often copyrighted audio and video files. The 
Recording Industry Association of America (RIAA) as a representative
of the U.S.\ recording industry investigates the illegal distribution
of sound recordings. The RIAA tries to identify the file sharing
network nodes distributing large amount of illegal content in order to
shut them down (ie. removing them from the network).

The motivation of the present work is to find out whether the
nodes can apply a simple local strategy to keep the peer-to-peer
network in a working state, or by continuously removing the most active
nodes from the network, the attacker can effectively prevent the
network from fulfilling its function. By local strategy we mean an
algorithm which uses only a constant amount of information, regardless
of the size of the network. Although our motivation was
specific, we believe that our model and results are more general and
widely applicable to various other networks as well.


The outline of this report is as follows. In Section~\ref{sec:related}
we very briefly discuss the literature closely related to the project.
In Section~\ref{sec:model} we define our network model, and justify
the model's assumptions. In Section~\ref{sec:results} we present three
different defensive strategies, and their performance
analysis. Finally, in Section~\ref{sec:conclusions} we give our conclusions.

\section{Related Work}
\label{sec:related}


There has been some work done on measuring the resilience of
different network structures, especially resilience of scale-free
networks to random failure and targeted attack \citep{albert00}. Here
failure means removing nodes randomly from the network, and targeted
attack means removing the high degree nodes from the network. Albert
and her co-workers showed that scale-free networks are not vulnerable
to random failure but very vulnerable to targeted attacks. A similar
approach was applied by \citet{broder00}.
\citet{callaway00} studied percolation, which is closely related to
network resilience, on random graphs with arbitrary degree
distributions. They used generating function methods to solve bond and
site percolation problems, in which the occupation probability was
a given function of the vertex degree. 

\section{Model Description}
\label{sec:model}

\subsection{How to attack a network?}

A network attack, ie. node removal from the network, can cause various
levels of damage to the structure (and function also) of the
network. In the classic work on network tolerance \citet{albert00}
studied how the diameter of the network changes after 1) removing
nodes randomly and 2) removing preferentially higher degree nodes. The
intuitive result was that scale-free networks are very resilient to
random node removal (ie. a failure) and not at all resilient to
preferential node removal (ie. a targeted attack).

In our model we assume that the attacker tries to inflict as much damage
as possible and removes the higher degree nodes from the
network. These nodes are not very difficult to find, even if the
attacker doesn't know the structure of the whole network
\citep{cohen02}. 

Let us assume that the degree distribution of the network is given by
$p_k (k=0,\ldots,n-1)$ where $n$~is the number of nodes in the
network. Thus the probability that a randomly chosen node in the
network has degree $k$ is $p_k$. Now, if we choose a random neighbor of
the randomly chosen node (let us assume that it has one), the
probability that this second node has degree $k$ is proportional to
$kp_k$, since the more edges this second node has, the higher the
probability is that it will be selected. (Here we assume that there
are no correlations between node degrees in the network, the
assortativity coefficient \citep{newman02a} is zero.) The same method is
described by \citet{cohen02}, in a different context.

In our model we assume that the attacker uses this method to find the
nodes with high degree in the network.

\subsection{The model assumptions}


We consider only \emph{undirected networks.} In real peer-to-peer
networks, this is not always the case, most of them are 
directed. 
The reason for assuming mutual connections (undirectedness) is
simplicity. The defensive strategies are more difficult to define on
directed networks, and also keeping the number of edges constant
(another assumption, comes later) is more difficult.

The network attacks are carried out in the way described in the
previous section. A random neighbor of a randomly chosen node is
removed with all its edges from the network. If the randomly chosen
node has no neighbors at all, no node is removed.
This method tends to remove the hubs (ie.~high degree nodes) from the
network. 

We keep the \emph{number of nodes} in the network constant. The reason 
for this is partly practical: we wanted to study the long term behavior of
the model and didn't want the network to shrink to a very small
size. The other reason is that we assumed that there are new nodes
joining to the network. The rate of the node removal
and the rate of the new nodes joining are the same in the model, and
this result.

Not only the number of nodes, but also the \emph{number of edges} is kept
constant in the model. This assumption is made because in real
networks edges have costs associated with them. A fixed number of
edges means fixed cost for maintaining the network. We wanted the
network to repair itself without increasing this cost.

In order to keep the number of edges constant, after the attack the
same number of edges are added to the network as the number of edges
deleted with the attacked node. One edge is given to the newly added
node, it will have degree one, and connects to a random node in the
network. The remaining edges are given to the former neighbors of the
attacked node (we will call these \emph{affected nodes} in the
following). While adding the new edges, we don't check for self-loops
(a node connects to itself) or multiple edges, in some cases they are
even needed to keep the number of edges constant. We can look at
self-loops as spare edges for the nodes, they can be used later to
create a ``real'' connection. Multiple edges can be considered as
stronger edges between two nodes.

The different \emph{defensive strategies} for the network are defined as the
methods the new edges are added to the affected nodes. A defensive
strategy is an algorithm used by an affected node to decide which node
it will connect to instead of the lost (attacked) node. The
information a node has about the structure of the network is
considered to be part of the algorithm. Sometimes we
call these strategies {\it rewiring strategies}, because the affected nodes
rewire from the attacked node to another one.

This is a \emph{discrete time model.}
The time steps of the simulation is defined by the attacks, there is
one attack in each time step. Only the affected nodes are active in
the network (the newly added node can also be considered active), the
other nodes don't initiate or remove connections. The affected nodes
react to the attack by creating new edges.


There are various structural properties of networks considered to be
important for their function. For efficient information flow, the
characteristic path length or the diameter of the network should be
small; if the edges have large costs, the redundancy of the network
should be minimal, etc. In this study we've focused on a very basic
structural property: the \emph{connectedness} of the network. It is clear that
in order to function, a peer-to-peer network should be connected. 
(For efficient function of course usually other properties
are needed.) More precisely we've considered the size of the largest
connected component of the network to compare the rewiring strategies,
after a large number of attacks, when the network can be assumed to be
in a steady state.

\section{Results}
\label{sec:results}

\subsection{The random rewiring strategy}


Let us first consider a very simple (but promising) possible defensive
strategy for the nodes. Here we do \emph{not} mean a strategy which
is easy to implement, but one which is easy to handle in simulations
or even in analytical calculations.


The random rewiring strategy is indeed completely random. The affected
nodes rewire to a randomly chosen node in the network (including also
the newly added node).


It is known since \citet{erdos59} that a random
graph has a giant connected component with probability~1 if the
average node degree in the graph is greater than~1. We also know that
when the average node degree is~1 there is a phase transition in the
system: the probability of having a giant cluster jumps from
probability~0 to probability~1. 

These facts have strong implications for our model. So long as we can
keep the network as a random graph, we will have a large connected
component. In the model we keep both the number of nodes and the
number of edges constant, thus the average node degree is also
constant. So if we can be sure that the random rewiring strategy
keeps the network in the random graph state, and we start with a
random graph with an average degree high enough, the network will have
a giant connected component even after the attacks. Note that starting
with a random graph is a quite strong assumption which is usually not
true in real peer-to-peer networks which show power-law degree
distributions \citep{ripeanu02}.

\begin{figure}
\includegraphics[width=0.48\textwidth]{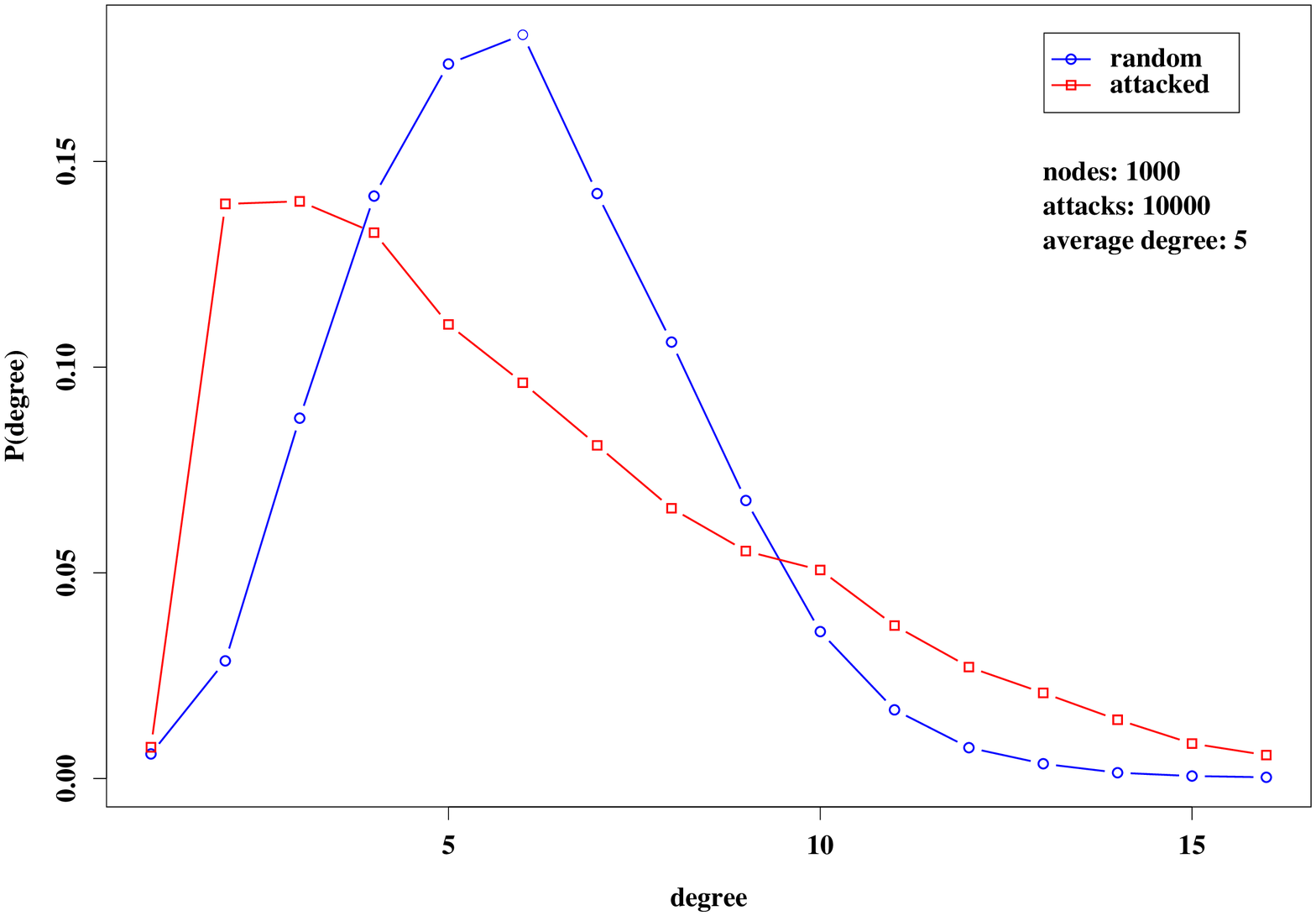}%
\caption{The degree
  distributions of a random graph (blue line, circles) and a 
  random graph after attacks and random rewiring repair (red
  line, squares). The plot shows that the random rewiring strategy
  does not keep a random graph random. This is because the attacks
  remove preferentially the nodes with high degree. }
\label{fig:rnd-att1}
\end{figure}

So we've investigated whether the attacks and the random rewiring
strategy keeps a random graph in the random state.
Fig.~\ref{fig:rnd-att1} shows that the degree distribution of a random
graph changes after the attacks and the random rewiring repair, the
graph is not a random graph any more. 

Even if the attacked network is not a random graph, it's degree
distribution is similar to a random graph so the random rewiring
strategy is still promising. Fig.~\ref{fig:rnd_greedy_time} indeed
shows that it performs very well.

\begin{figure}
\includegraphics[width=0.48\textwidth]{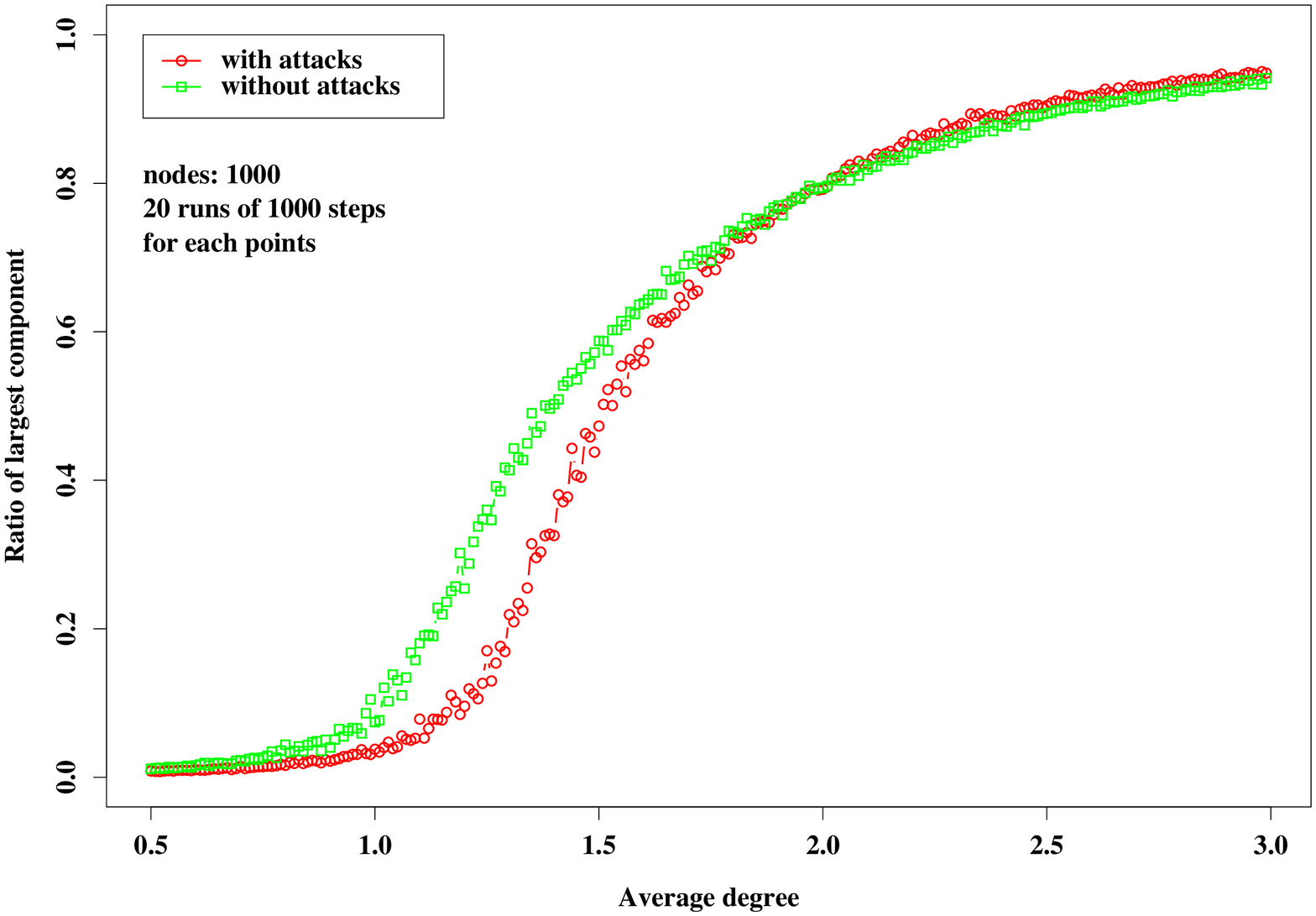}
\caption{Phase transitions in random graphs (green line, squares) and
  attacked random graphs (read line, circles). Further calculations
  are needed to prove the existence of the transition in the attacked
  graph. The transition threshold for the attacked network seems to be
  higher. } 
\label{fig:rnd-att2}
\end{figure}

As the structure of the random graph changes by using the random
rewiring strategy, it is an interesting question whether the giant
component's size in attacked graphs converges to a steady state and
if yes, whether there is a phase transition in the size of the largest
component in the attacked network controlled by the average
degree of the nodes. Although we don't give a proof for 
the existence of the phase transition, according to the experimental
results shown on Fig.~\ref{fig:rnd-att2} there is also a phase
transition in the attacked network, for which the phase transition
threshold seems to be different than the standard degree~1 threshold
in the random graphs. Further calculations are needed to calculate the
exact position of the phase transition.

The random rewiring strategy is efficient, however it is not plausible
in real  networks. In order to choose a node from the whole network
randomly, every single node has to know every other node in the
network. This means that the nodes have complete information (almost
complete, only the nodes have to be known, the edges don't) of the
whole network which is usually not true in real networks, especially
not in peer-to-peer networks. We also investigate other strategies,
which are implementable in practice.

\subsection{The greedy rewiring strategy}


The second rewiring strategy we've analyzed is a local
algorithm. Every node is considered having only local information (it
knows its neighbors), and behaving greedily, it tries to connect to a
good node in its neighborhood. A node is considered good if it has
many edges. 

More precisely the greedy strategy is defined as follows: the affected
node chooses a random neighbor and connects to the best
neighbor of it. The best neighbor is the neighbor with the highest
degree. If the affected node has no neighbors or second neighbors, the
node creates a self loop in order to keep the number of edges
constant. The affected nodes reconnect in random order.

\begin{figure}
\centering
\includegraphics[width=0.48\textwidth]{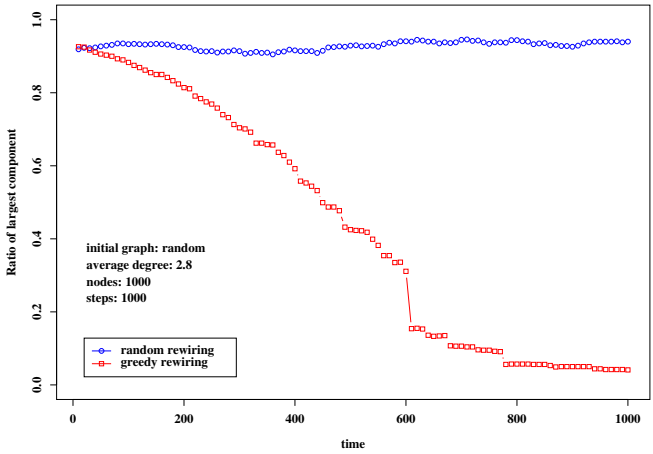}
\caption{The performance of the random (blue line, circles) and greedy
  (red line, squares) strategies 
  starting with a random graph. The plot shows the ratio of the number
  of nodes in the largest connected component to the network
  size. If the random rewiring is applied, 
  the network reserves the size of the largest component, even if the
  structure of  the initial random graph changes. The  
  greedy strategy always ends up with completely disconnected nodes
  after sufficiently many attacks.}
\label{fig:rnd_greedy_time}
\end{figure}

This strategy is clearly local and easy to implement in practice.
It is trivial however that it is not able to keep the network
connected. This is because this strategy cannot reconnect a network
which has fallen into two separate components as the result of an
attack. In the steady state the network consists almost exclusively of
isolated nodes. For the actual performance, see
Fig.~\ref{fig:rnd_greedy_time}.

\subsection{The `2nd neighbor' rewiring}


The third strategy is motivated by the failure of the local greedy and
the success of the global random strategy. This third strategy uses
more information than the greedy rewiring, but it is still intended to
be local.


The `2nd neighbor' rewiring involves keeping a list of all its second
neighbors by every node in the network. This list is used for an
affected node to rewire to a former neighbor of the attacked
node. After the rewiring, the lists of the nodes are updated.

\begin{figure}
\centering
\includegraphics[width=0.48\textwidth]{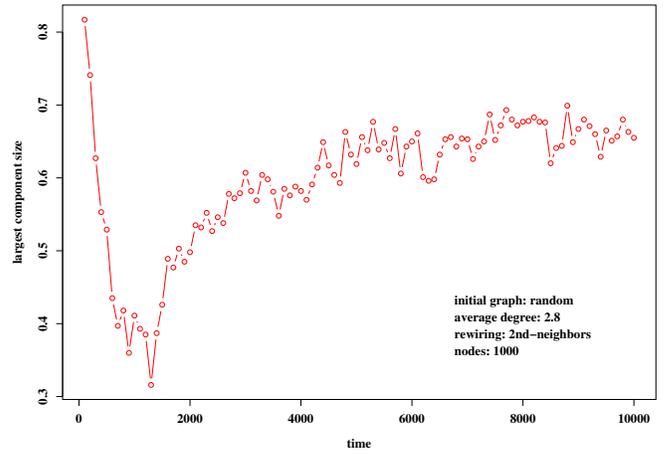}
\caption{The performance of the
  `2nd neighbor' strategy. As before, 
  the plot shows the changes in the ratio of the largest cluster in the
  network. In the short run, there is a steep decrease in the size of
  the largest component, but after that, the largest component starts
  to grow and reaches a quite high level. The 2nd neighbor strategy performs
  well in the long run. Also note the quite big
  fluctuations, even after the curve somewhat levels off.}
\label{fig:nei_perf1}
\end{figure}

The performance of the 2nd neighbor strategy is shown on
Fig.~\ref{fig:nei_perf1}. The plot shows that in the short run
this strategy performs quite poorly, there is a steep decrease in the
size of the giant component. In the long run however the performance
is much better, the size of the largest components starts to increase
and although it does not reach the original size in the starting
random graph and there are also quite big fluctuations in it, its
value is quite high.

According to Fig.~\ref{fig:nei_perf1} the performance of the
2nd neighbor strategy is adequate in the long run. However, it is unclear
why this strategy is able to keep the network connected. We do not know what
structural property changes in the graph in the decreasing regime,
what structural changes initiate the increase, and what kind of network
we get in the end. What we believe is the following. The 2nd neighbor
strategy restructures the network to be resilient even against
targeted attacks. In the first phase of the restructuring, the size of
the largest component decreases, but as soon as some structural
property is achieved, the largest component will be very resilient to
the attacks and remains connected. This strong large component is also
able to grow because the new nodes attach to it with significant
probability. Still there are some weak parts in this component, which
can be broken and separated by the attack, this causes the large
fluctuations. In the final phase the network reaches a state of
dynamic equilibrium, where new nodes are constantly attached to the
large connected component, but also some small parts are broken from
it. 

Further work is needed to investigate how the structural properties of
the network change as a result of the attacks and the rewiring.


\begin{figure}
\centering
\includegraphics[width=0.48\textwidth]{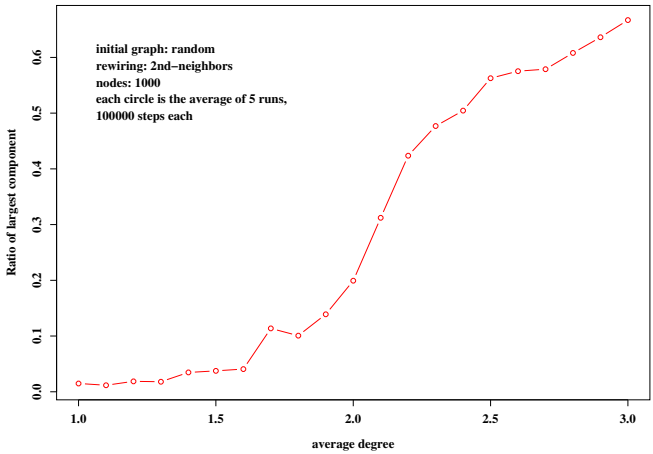}
\caption{The performance of the 2nd neighbor strategy in the function
  of the average node degree in the network. There might be a phase
  transition in the network, where an infinite giant component appears
  in infinite networks around average node degree~2 with
  probability~1.}  
\label{fig:nei_perf2}
\end{figure}

It is an interesting question if there is also a phase transition in
the networks resulted by the 2nd neighbor strategy by increasing the
average degree of the nodes. We've addressed this question, and made
experiments, for which the results are on Fig.~\ref{fig:nei_perf2}.
The results show that indeed there might be a phase transition in the
system around average degree~2, but more work is needed
to prove the existence of the phase transition. If this kind of
transition really exists in the system, that means that networks
having high enough average node degree and nodes applying the
2nd neighbor rewiring algorithm are very resilient to attacks.


Let us address the question of the amount of information required by
the 2nd neighbor rewiring strategy. Clearly, for a single node this is
proportional to the number of second neighbors it has. Ideally we want
every node to have a constant number of second neighbors,
independently of the size of the network; this is how a local strategy
was defined.

After some exploratory numerical simulations our impression is that
for networks without hubs the maximum number of second neighbors grows
much slower than the network size, probably as the logarithm of it.
For scale-free networks however, as they tend to contain big hubs, 
the required amount of information for a node is proportional to the
size of the whole network.

\section{Conclusions and Future Work}
\label{sec:conclusions}


In this report we have defined and analyzed a model for describing the
effect of attacks on peer-to-peer networks and studying the efficiency
of various defense mechanisms for the network. We've defined three
such mechanisms, the random, greedy and second neighbor rewiring, and
conclude that the last one offers fair performance on the long runs
and seems to be implementable in practice.


For future work, the most important task lies in quantifying the amount of
information needed by the nodes for the application of the 2nd
neighbor rewiring strategy in various networks. There are promising
results by \citet{newman01} using generating functions to give the
distribution in networks with arbitrary degree distributions. Also,
more numerical experiments should be conducted to find out how the
number of second neighbors scales with the network size for different
networks. 


It is also important to discover how the 2nd neighbor algorithm
restructures the network, what structural properties change during the
first (shrinking largest cluster) and second (growing largest cluster)
phase of the simulation.


While to have a connected network is required for the function of the
network, sometimes it is not enough and other structural properties
are also needed. It is important to examine how
the 2nd neighbor algorithm effects these, starting perhaps with the
characteristic path length of the network.

\section{Acknowledgement}


This work was done as a student project at the Santa Fe Institute
Complex Systems Summer School 2004. All the results presented in this
paper were achieved during the four weeks of the school. We thank the
Santa Fe Institute and the organizers of the summer school for their
support and help. We also thank the American Physical Society for the
\LaTeX{} style file used in this report.


G\'abor Cs\'ardi was partially funded by a joint grant of the National
Science Foundation and the Hungarian Academy of Sciences (grant number
0332075).

\bibliographystyle{abbrvnat}
\bibliography{papers}

\begin{thebibliography}{12}
\expandafter\ifx\csname natexlab\endcsname\relax\def\natexlab#1{#1}\fi
\expandafter\ifx\csname url\endcsname\relax
  \def\url#1{{\tt #1}}\fi

\bibitem[Albert and Barab{\'a}si(2002)]{albert02}
R.~Albert and A.-L. Barab{\'a}si.
\newblock Statistical mechanics of complex networks.
\newblock {\em Reviews of Modern Physics}, 74:\penalty0 47--97, 2002.

\bibitem[Albert et~al.(2000)Albert, Jeong, and Barab{\'a}si]{albert00}
R.~Albert, H.~Jeong, and A.-L. Barab{\'a}si.
\newblock Attack and error tolerance of complex networks.
\newblock {\em Nature}, 406:\penalty0 378--382, 2000.

\bibitem[Broder et~al.(2000)Broder, Kumar, Maghoul, Raghavan, Rajagopalan,
  Stata, Tomkins, and Wiener]{broder00}
A.~Broder, R.~Kumar, F.~Maghoul, P.~Raghavan, S.~Rajagopalan, R.~Stata,
  A.~Tomkins, and J.~Wiener.
\newblock Graph structure in the web.
\newblock {\em Computer Networks}, 33:\penalty0 309--320, 2000.

\bibitem[Callaway et~al.(2000)Callaway, Newman, Strogatz, and
  Watts]{callaway00}
D.~S. Callaway, M.~E.~J. Newman, S.~H. Strogatz, and D.~J. Watts.
\newblock Network robustness and fragility: Percolation on random graphs.
\newblock {\em Physical Review Letters}, 85:\penalty0 5468--5471, 2000.

\bibitem[Cohen et~al.(2003)Cohen, Havlin, and ben Avraham]{cohen02}
R.~Cohen, S.~Havlin, and D.~ben Avraham.
\newblock Efficient immunization strategies for computer networks and
  populations.
\newblock {\em Physical Review Letters}, 91:\penalty0 247901, 2003.

\bibitem[Dorogovtsev and Mendes(2002)]{dorogovtsev02}
S.~N. Dorogovtsev and J.~F.~F. Mendes.
\newblock Evolution of networks.
\newblock {\em Advances in Physics}, 51:\penalty0 1079--1187, 2002.

\bibitem[Erd{\H{o}}s and R{\'e}nyi(1959)]{erdos59}
P.~Erd{\H{o}}s and A.~R{\'e}nyi.
\newblock On random graphs.
\newblock {\em Publicationes Mathematicae}, 6:\penalty0 290--297, 1959.

\bibitem[Milojicic et~al.(2002)Milojicic, Kalogeraji, Lukose, Nagaraja, Pruyne,
  Richard, Rollins, and Xu]{milojicic02}
D.~S. Milojicic, V.~Kalogeraji, R.~Lukose, K.~Nagaraja, J.~Pruyne, B.~Richard,
  S.~Rollins, and Z.~Xu.
\newblock Peer-to-peer computing.
\newblock Technical Report HPL-2002-57, HP Laboratories, Palo Alto, 2002.

\bibitem[Newman(2002)]{newman02a}
M.~E.~J. Newman.
\newblock Assortative mixing in networks.
\newblock {\em Physical Review Letters}, 89:\penalty0 208701, 2002.

\bibitem[Newman(2003)]{newman03}
M.~E.~J. Newman.
\newblock The structure and function of complex networks.
\newblock {\em SIAM Review}, 45:\penalty0 167--256, 2003.

\bibitem[Newman et~al.(2001)Newman, Strogatz, and Watts]{newman01}
M.~E.~J. Newman, S.~H. Strogatz, and D.~J. Watts.
\newblock Random graphs with arbitrary degree distributions and their
  applications.
\newblock {\em Physical Review E}, 64:\penalty0 026118, 2001.

\bibitem[Ripeanu et~al.(2002)Ripeanu, Foster, and Iamnitchi]{ripeanu02}
M.~Ripeanu, I.~Foster, and A.~Iamnitchi.
\newblock Mapping the {G}nutella network: Properties of large-scale
  peer-to-peer systems and implications for system design.
\newblock {\em IEEE Internet Computing}, 6:\penalty0 50--57, 2002.

\end{thebibliography}

\end{document}